\documentclass[aps,prl,twocolumn,showpacs,floatfix,amsmath,amssymb,superscriptaddress]{revtex4}
\usepackage{graphicx}
\usepackage{float}
\usepackage{wasysym}
\usepackage[dvips]{color}
\usepackage{hyperref}
\usepackage{hypernat}
\usepackage{afterpage}
\usepackage{sidecap}
\usepackage{url}
\usepackage{longtable}
\usepackage{supertabular}
\def\etal{{\em{et al.}}}
\newcommand{\longoverbrace}[2]{\overbrace{#1}^{\text{\hbox to 0cm{\hss #2 \hss}}}}  
\newcommand{\longunderbrace}[2]{\underbrace{#1}_{\text{\hbox to 0cm{\hss #2 \hss}}}}

\begin{document}
\title{A Typical Medium Dynamical Cluster Approximation for the Study of Anderson Localization in Three Dimensions}

\author{C. E. Ekuma}
\altaffiliation{Electronic address: cekuma1@lsu.edu}
\affiliation{Department of Physics \& Astronomy, Louisiana State University,
Baton Rouge, Louisiana 70803, USA}
\affiliation{Center for Computation and Technology, Louisiana State University, Baton Rouge, Louisiana 70803, USA}

\author{H. Terletska}
\affiliation{Department of Physics \& Astronomy, Louisiana State University,
Baton Rouge, Louisiana 70803, USA}
\affiliation{Brookhaven National Laboratory, Upton, New York 11973, USA}

\author{K.-M. Tam}
\affiliation{Department of Physics \& Astronomy, Louisiana State University,
Baton Rouge, Louisiana 70803, USA}
\affiliation{Center for Computation and Technology, Louisiana State University, Baton Rouge, Louisiana 70803, USA}

\author{Z.-Y. Meng}
\affiliation{Department of Physics \& Astronomy, Louisiana State University,
Baton Rouge, Louisiana 70803, USA}
\affiliation{Center for Computation and Technology, Louisiana State University, Baton Rouge, Louisiana 70803, USA}
\affiliation{Department of Physics, University of Toronto, Toronto, Ontario M5S 1A7, Canada}

\author{J. Moreno}
\affiliation{Department of Physics \& Astronomy, Louisiana State University,
Baton Rouge, Louisiana 70803, USA}
\affiliation{Center for Computation and Technology, Louisiana State University, Baton Rouge, Louisiana 70803, USA}

\author{M. Jarrell}
\altaffiliation{Electronic address: jarrellphysics@gmail.com}
\affiliation{Department of Physics \& Astronomy, Louisiana State University,
Baton Rouge, Louisiana 70803, USA}
\affiliation{Center for Computation and Technology, Louisiana State University, Baton Rouge, Louisiana 70803, USA}

\begin{abstract}
\noindent We develop a systematic typical medium dynamical cluster approximation that provides a proper 
description of the Anderson localization transition in three dimensions (3D). Our method successfully 
captures the localization phenomenon both in the low and large disorder regimes, and allows us to study the 
localization in different momenta cells, which renders the discovery that the Anderson localization 
transition occurs in a cell-selective fashion. As a function of cluster size, our method systematically 
recovers the re-entrance behavior of the mobility edge and obtains the correct critical disorder strength 
for Anderson localization in 3D.
\end{abstract}

\pacs{72.15.Rn,72.80.Ng,02.70.Uu,64.70.Tg}

\maketitle 
\textit{Introduction}.-- The search for new methods to better understand Anderson 
localization~\cite{Anderson,PhysRevLett.42.673} 
remains an active area in the study of disordered electronic systems~\cite{50years}. Here, the scattering of 
charge carriers off random impurities~\cite{Bulka85,Lee} may inhibit their propagation across the sample 
leading to a phenomenon known as Anderson localization~\cite{Anderson}. Despite intensive studies, a proper 
mean-field theory of this phenomenon remains elusive. 

The most commonly used mean-field theory to study disordered systems is the coherent potential approximation 
(CPA)~\cite{p_soven_67,Velicky68}, where the original disordered lattice is replaced by an impurity embedded 
in an effective medium. The CPA successfully describes some one-particle properties, such as the density of 
states (DOS) in substitutional disordered alloys~\cite{p_soven_67, Velicky68}, but fails to capture the Anderson 
localization transition (ALT).  As a local approximation, the CPA is unable to capture crucial multiple 
backscattering interference effects that can lead to localization.  Cluster extensions of 
the CPA such as the dynamical cluster approximation (DCA)~\cite{Hettler2000,PhysRevB.63.125102} and the molecular 
CPA~\cite{MCPA} incorporate non-local effects; however, they still fail to describe the ALT. 
The average DOS calculated within such mean-field theories cannot distinguish between extended and localized 
states and it is not critical at the transition~\cite{PhysRevB.63.125102}; hence, it cannot be used as an order 
parameter. Finding a proper single-particle order parameter for the ALT capable of distinguishing between 
localized and extended states is a major challenge in the study of disordered electronic systems.   

While at the ALT the average DOS is not critical~\cite{Thouless}, the geometrical mean of the local DOS 
(LDOS)~\cite{Janssen98,Crow1988,Vollhardt,Derrida198429}, which better approximates 
the typical value of the LDOS, is actually critical. 
Dobrosavljevi\'{c} \etal~\cite{Vlad2003} incorporated such geometric averaging over disorder in the Typical 
Medium Theory (TMT) where the typical and not the average LDOS is used in the CPA self-consistency loop. They 
showed that the typical DOS (TDOS) obtained from geometric averaging over disorder becomes  critical at the 
transition, and hence can serve as an appropriate order parameter for the ALT.

The local TMT reproduces some of the expected features of the ALT, but fails to provide a proper description 
of the critical behavior in 3D.  It underestimates the critical disorder strength with $W_c^{TMT}\approx1.65$ instead of 
the numerical value $W_c\approx 2.1$~\cite{Slevin99,PhysRevB.63.045108,RevModPhys.80.1355,PhysRevB.84.134209} 
(in a unit where $4t=1$), and the critical exponent of the order parameter $\beta^{TMT}\approx1.0$ whereas
the recently reported value is $\beta\approx1.67$ ~\cite{PhysRevLett.105.046403,PhysRevB.84.134209}. 
Another crucial drawback of 
the local TMT in 3D is that it cannot describe the re-entrant behavior of the mobility edge (the energy separating 
extended and localized electron states) as seen in transfer matrix method studies~\cite{Bulka85,Fehske}. 
Hence, by its construction the 
TMT is able to describe the effects of strong localization due to disorder, but all non-local spatial correlation 
effects are missed~\cite{Byczuk2010}.  

A natural way to improve upon the local TMT is to construct its cluster extension using the DCA scheme, which 
systematically incorporates non-local effects. Recently, we extended the local TMT to a cluster version called 
the Cluster Typical Medium Theory (CTMT)~\cite{Ekuma-Arxiv}.  Here, the diagonal cluster-momentum-resolved density of 
states  is replaced by its typical value $\rho^c(K,\omega) = \exp(\langle \ln \rho^c(K,\omega)\rangle)$.  This scheme 
works well in lower dimensions, where weak localization effects are most pronounced, and our results reveal that all 
the states are localized in the large cluster limit.  However, this formalism does not properly describe the ALT in
a 3D lattice.  The reason is that in 3D, at a given disorder strength below the critical value $W_c$, there are 
regions of the DOS consisting of only localized states, and others only extended states. To capture this mixing of 
localized and extended states requires that different energy scales are treated separately. 
Our CTMT formalism fails in 3D because the DOS at each cluster site is first averaged over the cluster 
to obtain $\rho^c(K,\omega)$.  For large clusters, it will not contain any information about the localization
edge, and a theory based upon it is unable to distinguish between states above and below the localization edge.

In this Rapid Communication, to avoid such self-averaging issues in the TDOS, we propose a different Typical Medium 
DCA (TMDCA) method by explicitly separating out the local part of the TDOS and treating it with 
a geometric averaging over disorder configurations. In this way, we are able to obtain a proper TDOS 
that characterizes the ALT in 3D. The method we develop is a 
systematic self-consistent effective medium theory to study ALT in 3D, which (i) recovers the original 
local TMT scheme at N$_c$ = 1;  (ii) recovers the DCA for small $W$ 
(when all states are metallic); (iii) provides a proper way to \textit{treat the different energy scales} 
such that the \textit{characteristic re-entrant behavior of the mobility edge} is obtained;
(iv) captures the critical behavior 
of the ALT with \textit{correct critical disorder strength}; (v) provides a correct 
description of the Anderson insulator at large $W$ when all states are localized; 
and (vi) fulfills all \textit{the essential requirements expected of a ``successful'' cluster 
theory} ~\cite{a_gonis_92,PhysRevB.63.125102} including causality and translational invariance.

\textit{Method}.--
We consider the Anderson model of noninteracting electrons subjected to a random potential. The 
Hamiltonian is given by
\begin{equation} \label{eqn:model}
H=-\sum_{\langle i j \rangle}t_{ij}(c_{i}^{\dagger}c_{j}+h.c.)+\sum_{i}V_i n_{i}.
\end{equation}
The disorder is modeled by a local potential $V_i$ randomly distributed according to a probability 
distribution $P(V_i)$.  The operator $c_{i}^\dagger$($c_{i}$) creates (annihilates) an electron
on site $i$, $n_{i} = c_{i}^\dagger c_{i}$ is the number operator, and $t_{ij}$ is the hopping matrix 
element between nearest-neighbor (NN) sites $\langle i,j\rangle$.  We set $4t = 1$ as the energy unit, 
and use a ``box'' distribution with $P(V_i)=\frac{1}{2W}\Theta(W-|V_i|)$, where $\Theta(x)$ is a  step 
function. We use the short-hand notation: $\langle...\rangle=\int dV_i P(V_i) (...)$ for disorder averaging. 

To solve the Hamiltonian (\ref{eqn:model}) we utilize a modification of the standard DCA 
procedure~\cite{PhysRevB.63.125102}. Here, the original lattice model is mapped onto 
a periodic cluster of size N${}_c$ = L$_c^3$ embedded in a self-consistent typical 
medium characterized by a non-local hybridization function $\Gamma(K,\omega)$. Hence, 
spatial correlations up to a range $\xi \lesssim L_c$    
are treated explicitly, while the longer length scale physics is described at the mean-field level. 
The mapping is accomplished by dividing the first Brillouin zone into $N_c$ non-overlapping cells 
of equal size.  The lattice Green function is coarse-grained over the cells, and 
the cluster self-energy is subtracted to form the cluster-excluded Green function 
${\cal G}(K,\omega)=(\omega-\Gamma(K,\omega)-\bar{\epsilon}_K)^{-1}$, where $\bar{\epsilon}_K$ is 
the coarse-grained bare dispersion.  ${\cal G}(K,\omega)$ is Fourier transformed to form the real 
space ${\cal G}_{n,m} = \sum_{K} {\cal G}(K)\exp(i K\cdot(r_n-r_m))$.  Then for each randomly
chosen disorder configuration $V$, we calculate the cluster Green function $G^c(V)=({\cal G}^{-1}-V)^{-1}$.

From this quantity we obtain the typical density of states $\rho_{typ}^c(K,\omega)$ which 
is constructed as 
\begin{multline} \label{Eq:rho_typ_definition}
\rho_{typ}^c(K,\omega)
=
\longoverbrace{\exp\left(\frac{1}{N_c} \sum_{i=1}^{N_c} \left\langle \ln \rho_{i}^c (\omega,V)  \right\rangle\right)}{local TDOS} \times \\ 
\longunderbrace{\left\langle \frac{\rho^c(K,\omega,V)}{\frac{1}{N_c} \sum_{i} \rho_{i}^c (\omega,V)} \right\rangle }{non-local}.
\end{multline}
Here, $\rho_i^c(\omega,V)=-\frac{1}{\pi}\textnormal{Im}G^c_{ii}(\omega,V)$ while 
$\rho^c(K,\omega,V)=-\frac{1}{\pi}\textnormal{Im}G^c(K,\omega,V)$ is obtained from the diagonal
Fourier transform of the cluster Green function $G^c_{ij}(\omega,V)$.

As mentioned in the Introduction, to avoid self-averaging as N${}_c$ increases, we modify our 
CTMT~\cite{Ekuma-Arxiv} scheme in the way we calculate the spectra $\rho_{typ}^c(K,\omega)$ used in 
the self-consistency. In particular, as shown in Eq.~\ref{Eq:rho_typ_definition}, we separate the  
``local TDOS'', and treat it with geometrical averaging over disorder, from the ``non-local'' part
which is treated via algebraic averaging.

This $\rho_{typ}^{c}(K,\omega)$ possesses the following properties:
for N${}_c = 1$, it reduces to the local TMT with $\rho_{typ}^{c}(\omega)=\exp \langle \ln\rho^c(\omega,V)\rangle$. 
At low disorder strength, $W \ll W_c$, the local real space prefactor
$\left< \ln\rho_{i}^c(\omega,V)\right> \approx \ln \left<\rho_{i}^c(\omega,V) \right>$.
Then $\rho_{typ}^{c}(K,\omega)$ reduces to the DOS calculated in DCA scheme, with
$
\rho_{typ}^{c}(K,\omega)\rightarrow\left< \rho^c(K,\omega,V)\right>.
$

From Eq.~\ref{Eq:rho_typ_definition}, the disorder averaged typical cluster Green function is obtained 
using the Hilbert transform 
$ G_{typ}^c(K,\omega)=\int d \omega' \displaystyle \frac{\rho_{typ}^c(K,\omega')}{\omega - \omega'}$.
Finally, the self-consistency loop is closed by calculating the coarse-grained cluster Green function
of the lattice 

$
\overline{G} (K,\omega)  
= \int \displaystyle \frac{N^c_0(K,\epsilon) d\epsilon}{(G^c_{typ} (K,\omega))^{-1} + \Gamma (K,\omega) - 
 \epsilon + \overline{\epsilon}(K) +\mu},
$
where $N^c_0(K,\epsilon)$ is the bare partial density of states.

We note that our formalism  preserves causality just as the DCA~\cite{PhysRevB.63.125102}, 
since all the Green functions are causal, both the DOS and the TDOS calculated from them are 
positive definite.  Also, we observe that as N${}_c$ increases, our method systematically 
interpolates between the local TMT and the exact result. 

\begin{figure}[h!]
 \includegraphics[trim = 0mm 0mm 0mm 0mm,width=1\columnwidth,clip=true]{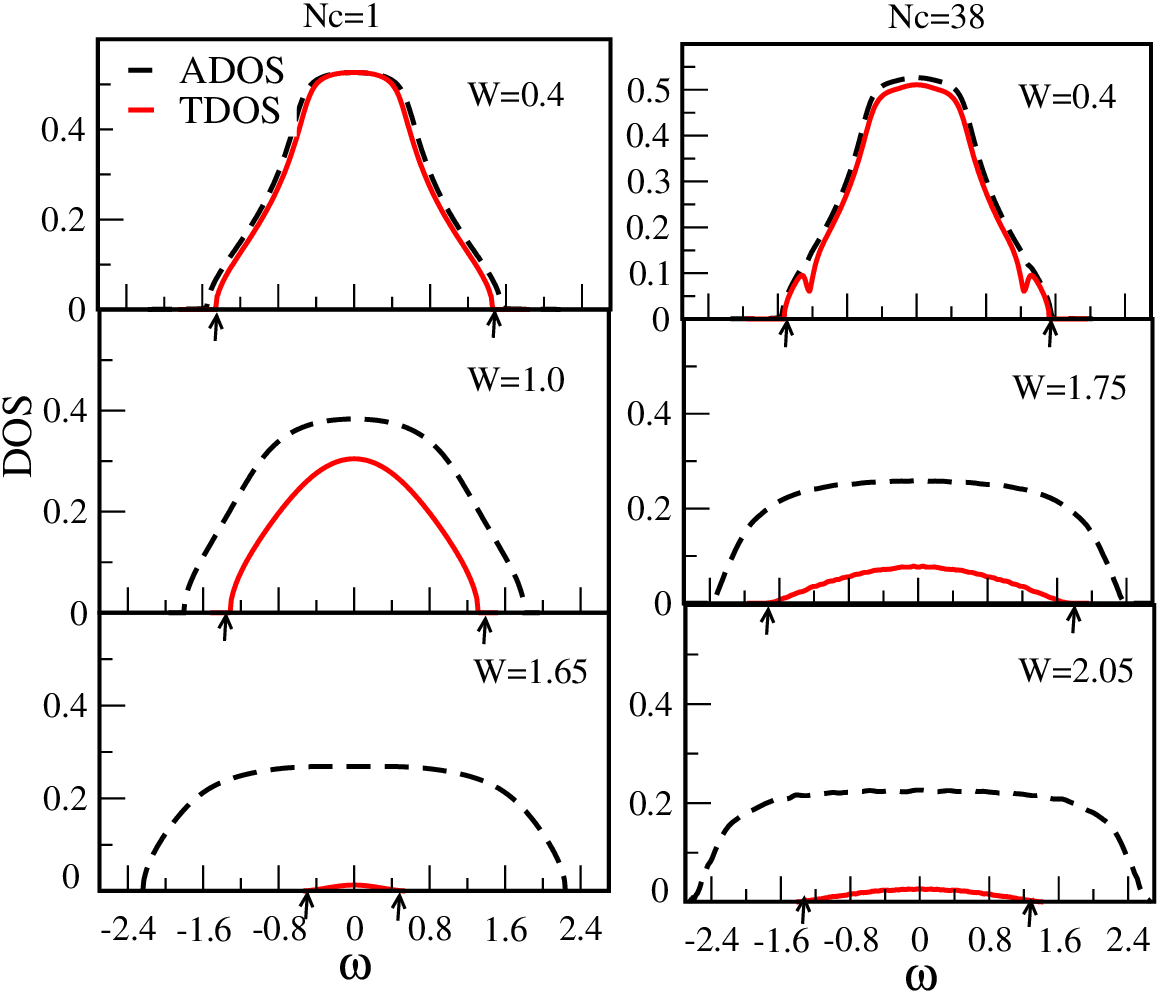}
\caption{(Color online). Evolution of the ADOS and TDOS at different
disorder strength  $W$ 
for the TMT (N${}_c$ = 1) and TMDCA with N${}_c = 38$. At low disorder, 
where all the states are metallic, the TDOS is the same as the ADOS. 
As $W$ increases the TDOS gets suppressed. In the local TMT, 
the mobility edge (indicated by arrows) moves towards $\omega=0$ monotonically. 
In the TMDCA the mobility edge 
first moves to higher energy, and around $W > 1.75$ it starts moving towards the band 
center, indicating that TMDCA can successfully capture the re-entrance behavior
missing in the TMT scheme.
} 
\label{Fig1:tdos_ados}
\end{figure}

\textit{Results and Discussion}-- 
We start the discussion of our results by comparing the algebraically averaged DOS 
(ADOS) and the TDOS for N${}_c$ = 1 and 38 at 
various disorder strengths (Fig.~\ref{Fig1:tdos_ados}).  Our TMDCA scheme for 
N${}_c$ = 1 corresponds to the original TMT procedure.  The ADOS is obtained from the conventional 
DCA scheme, where the ADOS is used in the self-consistency. 
As can be seen from Fig.~\ref{Fig1:tdos_ados} for both TMT (N${}_c$ = 1) and 
TMDCA (N${}_c$ = 38), the ADOS remains finite while the TDOS gets suppressed as $W_c$ 
is approached.

Hence, the TDOS indeed serves as a proper order parameter of the ALT.  In addition, at low disorder, 
$W = 0.4$, for N${}_c$ = 38, the ADOS and TDOS are practically the same, indicating that our TMDCA 
procedure at $W \ll W_c$ reduces to the DCA scheme in agreement with our analytical analysis described 
above.  Moreover, a crucial difference between the local TMT at N${}_c = 1$ and TMDCA at 
N${}_c = 38$ can be seen from the comparison of left and right panels of Fig.~\ref{Fig1:tdos_ados}. 
The mobility edge, separating extended from localized states, is defined by the boundary 
of the TDOS and indicated by arrows. For the local TMT the edge always gets narrower with increasing
$W$, while for TMDCA, the mobility edge first expands and then retracts, hence giving 
rise to the re-entrance behavior, which is missed in the local TMT.

\begin{figure}[b!]
\includegraphics[trim = 0mm 0mm 0mm 0mm,width=1\columnwidth,clip=true]{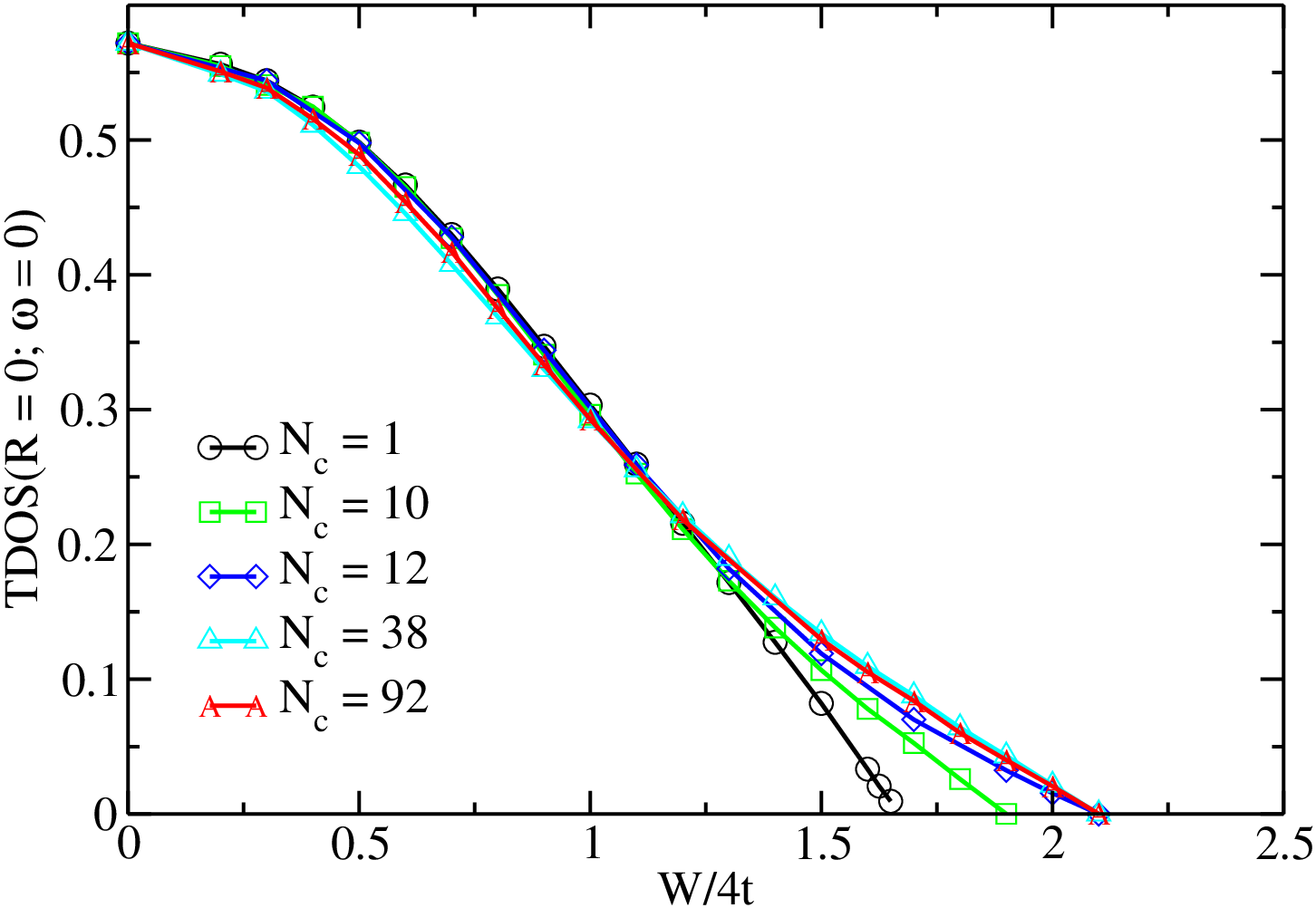}
\caption{(Color online). The TDOS\,$(\omega=0)$ vs.\  $W$ for different cluster 
sizes N${}_c=1,10,12,38,92$. The TDOS\,$(\omega=0)$ vanishes at $W_c$ where all states become localized.
For N${}_c=1$ (TMT), the critical disorder strength $W_c^{N_c=1} \approx 1.65$. As N${}_c$ increases, 
$W_c$ increases quickly to $W_c^{N_c\geq12}\approx 2.10 \pm 0.01$.}
\label{Fig2:tdos_vs_V}
\end{figure}
Next, we consider the evolution of $W_c$ with N${}_c$.  Figure~\ref{Fig2:tdos_vs_V} shows the TDOS 
at the band center as a function of $W$ for several N${}_c$. $W_c$ is defined by the vanishing of the 
TDOS\,($\omega=0$). Our results show that as cluster size N${}_c$ increases, for N${}_c\ge12$ the 
$W_c$ systematically increases until it converges 
to $W_c\approx 2.1$ (for details, see Supplemental Material (SM)~\cite{TMDCA_Supp}) which is in 
good agreement with the values reported in the literature
~\cite{Slevin99,Bulka85,Fehske,PhysRevB.76.045105,PhysRevB.63.045108,PhysRevB.84.134209,PhysRevLett.105.046403}.
This cluster is the first one with a complete NN shell (for details, see 
SM~\cite{TMDCA_Supp}). From this cluster onward, $W_c$ converges to $\approx$ 2.1. 
Fitting the data for the two largest clusters starting from $W \approx 1.0$ 
with a power law TDOS$(\omega=0)=a_0|W-C|^\beta$, 
we obtain $\beta >$ 1.40 which is greater than a single site TMT value 
of $\beta^{TMT} = 1.0$; but it is still smaller than the most recently
reported $\beta\approx 1.67$~\cite{PhysRevLett.105.046403,PhysRevB.84.134209,Beta_nu}. We note that other mean-field 
methods reported $\beta$ $\lesssim$ 1.0~\cite{PhysRevLett.48.699}.  
In our method, we note that it is unlikely that we can calculate the critical disorder strength and 
exponents as precisely as diagonalization and transfer matrix methods~\cite{Slevin99,Bulka85,Fehske,PhysRevB.76.045105,
PhysRevB.63.045108,PhysRevB.84.134209,PhysRevLett.105.046403}. However, the advantage of our 
method is that we can incorporate both interactions and realistic electronic structure as in, e.g., the 
dynamical mean-field theory~\cite{RevModPhys.68.13} and other DCA calculations 
(see, e.g., Ref.~\cite{PhysRevB.73.085106}).

The probability distribution function (PDF) is a natural way to characterize the 3D Anderson localization 
transition. This is due to the fact 
that the ``typical'' value of a ``random'' variable corresponds to the most probable value of the 
PDF~\cite{Ekuma-Arxiv,PhysRevLett.105.046403}.

\begin{figure}[t!]
 \includegraphics[trim = 0mm 0mm 0mm 0mm,width=1\columnwidth,clip=true]{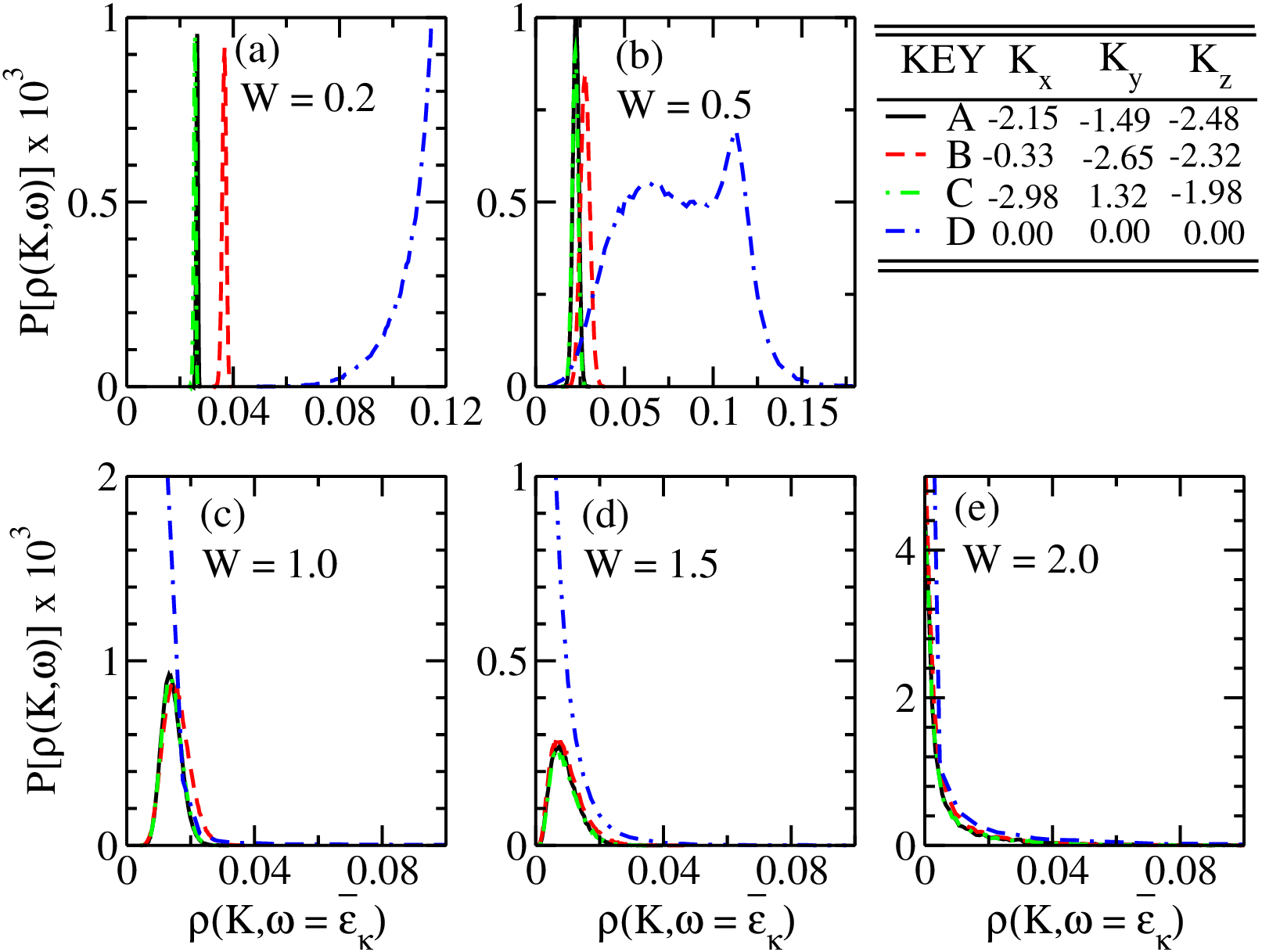}
\caption{(Color online). The evolution of the probability distribution of the cluster-momentum-resolved 
DOS at different cluster cells, PDF[$\rho(K,\omega=\bar{\epsilon}_K)$],  with increasing $W$ 
for N${}_c$ = 38. The labels A--D and their associated momenta $K$ correspond to  each of the 
four distinct cells obtained using the point-group and particle-hole symmetry 
($\rho(K,\omega) = \rho(Q-K,-\omega)$, with $Q= (\pi,\pi,\pi)$) of the cluster. Before the 
localization transition, the edge cells (corresponding to (0,0,0) and ($\pi,\pi,\pi$) label 
D)  develop a log-normal distribution while other cells remain Gaussian at small and moderate 
$W$ ((a)-(d)).  Close to the critical disorder strength $W = 2.0$, panel (e), all the cells 
show log-normal distributions.    
} 
\label{Fig3: histogram}
\end{figure}

Since, for a proper description of electron localization in disordered systems, one should focus on the distribution 
functions of the quantities of interest~\cite{Anderson}, we calculate the PDF of the cluster-momentum-resolved 
DOS $\rho(K,\omega=\bar{\epsilon}_K)$ (at different momenta cells $K$ and energy $\omega=\bar{\epsilon}_K$)
sampled over a large number of disorder configurations.  Our results for the evolution of the 
PDF[$\rho(K,\omega=\bar{\epsilon}_K)$] with $W$ are shown in Fig.~\ref{Fig3: histogram}.

From statistical studies of disordered systems~\cite{Janssen98, Fehske}, it is known that 
for extended states, when the amplitude of the wave function is approximately the same on every site, 
the distribution of the local DOS is Gaussian and the most probable value coincides with the arithmetic 
mean. On the other hand, for localized states, which have substantial weight on a few sites only, the 
distributions develop long tails and are extremely asymmetric with a log-normal shape.  Most of the weight 
is concentrated around zero, and its most probable value is much smaller than the arithmetic mean. 
As can be seen from Fig.~\ref{Fig3: histogram}, we indeed observe such behavior in our results.
In particular, we find that as $W$ increases, different cluster cells localize at different rates. The 
cells centered at cluster momenta $K = (0,0,0)$ and $(\pi,\pi,\pi)$ (labeled as D in Fig.~\ref{Fig3: histogram}) 
and energy at the band edges have the same PDF. They localize much faster than other cells 
with lower energies. The PDFs of these edge cells exhibit log-normal distribution far earlier than other 
cells which remain Gaussian up to moderate disorder strengths $W\sim$ 1.0 [panel (c)].  However, close to 
$W_c$ [cf. Fig.~\ref{Fig3: histogram} (e)], all the cells show log-normal distributions with their most 
probable values peaked close to zero. For $W=W_c$ all the states are localized and the system 
undergoes a full ALT in agreement with numerically exact results~\cite{PhysRevB.81.155106}. Hence, the localization 
transition occurs as a ``\textit{momentum cell-selective Anderson localization transition}''.

\begin{figure}[htb!]
 \includegraphics[trim = 0mm 0mm 0mm 0mm,width=1\columnwidth,clip=true]{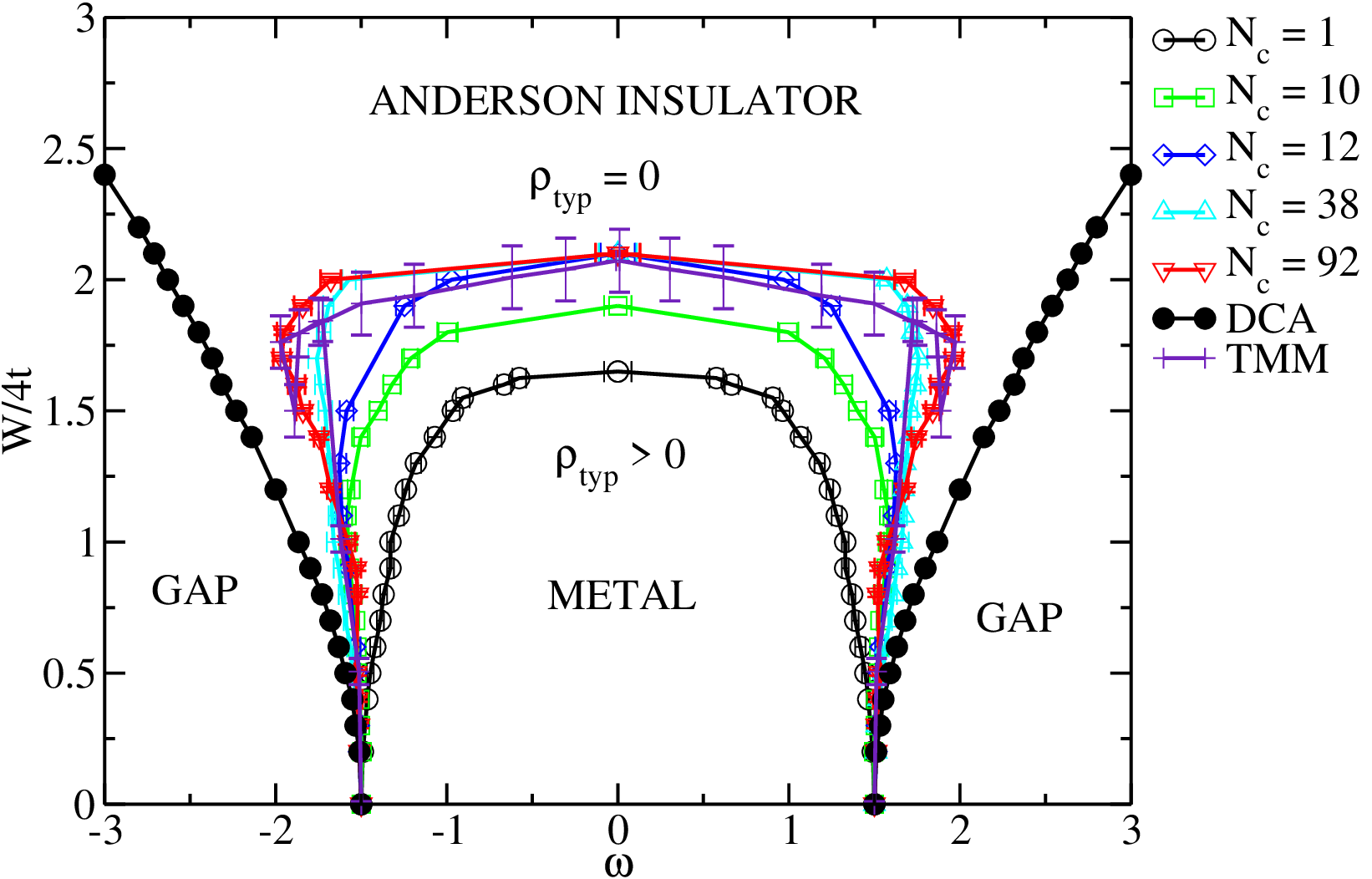}
\caption{(Color online). Phase diagram of the Anderson localization transition in 3D obtained 
from TMDCA simulations. As N$_c$ increases, a systematic improvement of the trajectory of the 
mobility edge is achieved. At large enough N$_c$ and within computation error, our results 
converge to those determined by the TMM~\cite{Bulka85}.}
\label{Fig4: phase diagram}
\end{figure}

Finally, in Fig.~\ref{Fig4: phase diagram} we present the phase diagram in the disorder-energy (W-$\omega$) 
plane for the 3D ALT constructed from our TMDCA procedure.  Here, we show the mobility edge trajectories 
given by the frequencies where the TDOS vanishes at a given disorder strength $W$, and the band edge 
determined by the vanishing of the ADOS calculated within the DCA.
For comparison, we also present the numerical results from the transfer matrix 
method in Ref.~\cite{Bulka85}. For large enough clusters, N${}_c = 92$, our results converge 
to the their results within the errors of both approaches. In particular, as N${}_c$ increases, the 
mobility edge trajectories are systematically reproduced, with re-entrance behavior gradually 
captured.

As evident from Fig.~\ref{Fig4: phase diagram}, at N${}_c$ = 12 (the first cluster with 
complete NN shell), the $W_c$ at $\omega=0$ quickly converges to $W_c=2.1$, while, the trajectory 
of the mobility edge continues to change with N${}_c$. This may
be understood from the different localization mechanisms for states at the band center and edge~\cite{Bulka85}.
States at the band center become localized mainly due to coherent backscattering, while those above 
and below the bare band edges are initially localized in deeply trapped states.  They become delocalized 
with increasing $W$ due to the increasing DOS and hence increasing quantum tunneling between the deeply 
trapped states.  They finally become localized again with increasing disorder, which explains the 
re-entrant behavior. Since coherent backscattering requires a retracing of the electronic path,
the effective length scales captured by the cluster are doubled, so $W_c$ converges very quickly
at the band center. On the other hand, the quantum tunneling mechanism has no path doubling and 
requires multiple deeply trapped states on the cluster and therefore converges more
slowly with N${}_c$.

\textit{Conclusions}--
In this Rapid Communication we develop a DCA-based typical medium theory (TMDCA) to study 
Anderson localization in three dimensions. 
The developed TMDCA presents a successful self-consistent, causal, and numerically 
manageable effective medium scheme of Anderson localization.
Employing the one-particle typical density of states as an order parameter of the 
Anderson transition, the TMDCA gives the critical disorder strength of 
$W_c=2.10 \pm 0.01$  which is in very good
agreement with the acceptable value in literature, and it is of noticable improvement 
over the single-site TMT result of $W_c=1.65$.
Moreover, our method systematically captures the re-entrance behavior of the mobility edge 
trajectories, which are absent in the local TMT scheme. Our analyses further shows \textit{a cell-selective 
Anderson localization transition}, 
with different cluster cells localizing at different rates. Our TMDCA method is easy to implement and computationally 
inexpensive since it requires only the computer time required to diagonalize small clusters, average over the disorder, 
and iterate to convergence. Once combined with electronic structure calculations\cite{PhysRevB.41.9701} and more 
sophisticated many-body techniques for electron interactions, it will open a new avenue for studying localization 
phenomenon in real materials as well as the competition between disorder and electron correlations.

\textit{Acknowledgments}--
We thank V.\ Dobrosavljevi\'{c}, S.-X.\ Yang, and C.\ Moore for useful discussions. This work is 
supported by the DOE BES CMCSN Grant No. DE-AC02-98CH10886 (H.T.) and SciDAC Grant No. DE-SC0005274 (M.J.\ and K.M.T), and the NSF 
EPSCoR Cooperative Agreement No.\ EPS-1003897 (C.E., Z.Y.M., and J.M.). Supercomputer support is 
provided by the Louisiana Optical Network Initiative (LONI) and HPC@LSU computing resources.


\begin{thebibliography}{99}
\bibitem{Beta_nu}
The scaling of the order parameter 
exponent~\cite{PhysRevB.84.134209,PhysRevLett.105.046403} 
is $\beta = (\alpha_o-d)\nu$, where $\nu$ is the correlation length exponent and $\alpha_o$ is the zero moment 
of the multifractal singularity strength. 
The most recent estimates as reported in Ref.~\cite{PhysRevB.84.134209} are 
$\alpha_o = 4.048$, $\nu=1.59$ and $\beta=1.67$.
\end{thebibliography}

\end{document}


\title{A Typical Medium Dynamical Cluster Approximation for the Study of Anderson Localization in Three Dimensions:
Supplementary Notes}

\author{C. E. Ekuma}
\altaffiliation{Electronic address: cekuma1@lsu.edu}
\affiliation{Department of Physics \& Astronomy, Louisiana State University,
Baton Rouge, Louisiana 70803, USA}
\affiliation{Center for Computation and Technology, Louisiana State University, Baton Rouge, Louisiana 70803, USA}

\author{H. Terletska}
\affiliation{Department of Physics \& Astronomy, Louisiana State University,
Baton Rouge, Louisiana 70803, USA}
\affiliation{Brookhaven National Laboratory, Upton, New York 11973, USA}

\author{K.-M. Tam}
\affiliation{Department of Physics \& Astronomy, Louisiana State University,
Baton Rouge, Louisiana 70803, USA}
\affiliation{Center for Computation and Technology, Louisiana State University, Baton Rouge, Louisiana 70803, USA}

\author{Z.-Y. Meng}
\affiliation{Department of Physics \& Astronomy, Louisiana State University,
Baton Rouge, Louisiana 70803, USA}
\affiliation{Center for Computation and Technology, Louisiana State University, Baton Rouge, Louisiana 70803, USA}
\affiliation{Department of Physics, University of Toronto, Toronto, Ontario M5S 1A7, Canada}

\author{J. Moreno}
\affiliation{Department of Physics \& Astronomy, Louisiana State University,
Baton Rouge, Louisiana 70803, USA}
\affiliation{Center for Computation and Technology, Louisiana State University, Baton Rouge, Louisiana 70803, USA}

\author{M. Jarrell}
\altaffiliation{Electronic address: jarrellphysics@gmail.com}
\affiliation{Department of Physics \& Astronomy, Louisiana State University,
Baton Rouge, Louisiana 70803, USA}
\affiliation{Center for Computation and Technology, Louisiana State University, Baton Rouge, Louisiana 70803, USA}

\pacs{72.15.Rn,72.80.Ng,02.70.Uu,64.70.Tg}

\maketitle

\section{Generation of Three-Dimensional Clusters}
We generate and utilize the cluster geometries following the grading scheme 
of Betts \etal~\cite{betts-3d}.  In Table~\ref{tab:clus3d} and Table~\ref{tab:clus3d_neighbor} we specify 
the cluster  geometries and other important parameters of the bipartite clusters used in our computations.
\begin{table}[htb] 
\centering
\caption{Three-dimensional cluster geometries of the best bipartite (B clusters), next best bipartite (C clusters), and 
next-next best bipartite (D clusters). The $a_{i}$ denote the cluster lattice vectors, IMP is the imperfection, 
C is the cubicity, and S is the number of complete shells.} 
\begin{tabular}{cccccccc}
\hline\hline
$N_c$ & $\vec{a}_1$ & $\vec{a}_2$ & $\vec{a}_3$ & IMP & C & S \\ 
\hline 
6B&(1, 0, 3)& (4, 1,-1)& (2, 0, 0)&2&1.014 & 0\\
6C&(1, 1, 2)& (4, 1,-1)& (2, 0, 0)&2&1.016 & 0\\
6D&(1, 0, 1)& (2, 1,-1)& (1, 2, 1)&2&1.019 & 0\\
10B&(1, 0, 3)& (4, 1,-3)& (-2,-1,-1)&1&1.005 & 0\\
10C&(1, 0, 3)& (3, 3, 4)& (1, 1,-2)&1&1.013 & 0\\
10D&(1, 1, 2)& (3, 2, 1)& (3,-1,-4)&1&1.018 & 0\\
12B&(1, 1, 2)& (4, 1,-3)& (3, 3, 2)&0&1.010 & 1\\
12C&(1, 1, 2)& (2,-4, 2)& (-2, 1,-1)&0&1.017 & 1\\
12D&(1, 1, 2)& (4,-2, 2)& (1,-2,-3)&0&1.018 & 1\\
14B&(1, 0, 3)& (2, 1, 1)& (1, 4,-3)&1&1.008 & 1\\
14C&(1, 1, 2)& (4,-1,-3)& (3, 2, 1)&1&1.011 & 1\\
14D&(1, 1, 2)& (2, 1,-1)& (1,-2, 1)&1&1.018 & 1\\
16B&(1, 1, 2)& (4, 1, 3)& (0, 2,-2)&2&1.011 & 1\\
16C&(1, 1, 2)& (2,-2, 0)& (1, 1,-2)&2&1.012 & 1\\
16D&(1, 1, 2)& (4, 1, 3)& (2, 1,-3)&2&1.012 & 1\\
38B&(1, 2, 3)& (3,-1,-2)& (2,-2, 2)&0&1.087 & 2\\
38C&(1, 2, 3)& (3,-1,-2)& (2, 3,-1)&0&1.117 & 2\\
38D&(1, 1, 4)& (3, 2, 1)& (2,-2, 2)&0&1.144 & 2\\
44B&(1, 2, 3)& (3, 2,-1)& (2,-2, 2)&3&1.036 & 2\\
44C&(1, 1, 4)& (3, 1,-2)& (2,-2, 2)&3&1.072 & 2\\
44D&(1, 2, 3)& (2,-2, 2)& (1, 4,-3)&3&1.076 & 2\\
80B&(1, 1, 4)& (3, 2,-3)& (3,-3, 2)&4&1.054 & 2\\
80C&(1, 1, 4)& (4,-2, 2)& (2, 3,-3)&4&1.068 & 2\\
80D&(1, 1, 4)& (4, 3, 1)& (3,-3, 2)&4&1.079 & 2\\
92B&(1, 3, 4)& (3,-2, 3)& (2, 4,-2)&2&1.085 & 3\\
92C&(1, 3, 4)& (4,-1,-3)& (3,-2, 3)&2&1.102 & 3\\
92D&(1, 3, 4)& (4,-1,-3)& (2, 4,-2)&2&1.119 & 3\\
\hline \hline
  \end{tabular}
  \label{tab:clus3d}
\end{table}

\begin{table}[h!]
\caption{Explicit description of the neighbors sites in three-dimensional clusters showing the shell number 
(with the nearest neighbor shell being 1, etc.), number of neighbors in that shell on the lattice (LS), and 
the number of neighbors in that
shell on the cluster (BS) for the various clusters in Table~\ref{tab:clus3d}. The smaller clusters with a full
nearest-neighbor shell have $N_c=12$, the smaller clusters with a complete next-nearest-neighbor shell have $N_c=38$, 
and with a complete next-next-nearest-neighbor shell have $N_c=92$.} 
\begin{tabular}{cccc|cccc|cccc}
\hline\hline
Nc & Shell & LS & BS & Nc & Shell & LS & BS & Nc & Shell & LS & BS \\ 
\hline
6B & 1 & 6 & 3 & 6C & 1 & 6 & 3 & 6D & 1 & 6 & 3 \\ 
 & 2 & 18 & 2 &  & 2 & 18 & 2 &  & 2 & 18 & 2 \\ 
10B & 1 & 6 & 5 & 10C & 1 & 6 & 5 & 10D & 1 & 6 & 5 \\ 
 & 2 & 18 & 4 &  & 2 & 18 & 4 &  & 2 & 18 & 4 \\ 
12B & 1 & 6 & 6 & 12C & 1 & 6 & 6 & 12D & 1 & 6 & 6 \\ 
 & 2 & 18 & 5 &  & 2 & 18 & 5 &  & 2 & 18 & 5 \\ 
14B & 1 & 6 & 6 & 14C & 1 & 6 & 6 & 14D & 1 & 6 & 6 \\ 
 & 2 & 18 & 6 &  & 2 & 18 & 6 &  & 2 & 18 & 6 \\ 
 & 3 & 38 & 1 &  & 3 & 38 & 1 &  & 3 & 38 & 1 \\ 
16B & 1 & 6 & 6 & 16C & 1 & 6 & 6 & 16D & 1 & 6 & 6 \\ 
 & 2 & 18 & 7 &  & 2 & 18 & 7 &  & 2 & 18 & 7 \\ 
 & 3 & 38 & 2 &  & 3 & 38 & 2 &  & 3 & 38 & 2 \\ 
38B & 1 & 6 & 6 & 38C & 1 & 6 & 6 & 38D & 1 & 6 & 6 \\ 
 & 2 & 18 & 18 &  & 2 & 18 & 18 &  & 2 & 18 & 18 \\ 
 & 3 & 38 & 13 &  & 3 & 38 & 13 &  & 3 & 38 & 13 \\ 
44B & 1 & 6 & 6 & 44C & 1 & 6 & 6 & 44D & 1 & 6 & 6 \\ 
 & 2 & 18 & 18 &  & 2 & 18 & 18 &  & 2 & 18 & 18 \\ 
 & 3 & 38 & 16 &  & 3 & 38 & 16 &  & 3 & 38 & 16 \\ 
 & 4 & 66 & 3 &  & 4 & 66 & 3 &  & 4 & 66 & 3 \\ 
80B & 1 & 6 & 6 & 80C & 1 & 6 & 6 & 80D & 1 & 6 & 6 \\ 
 & 2 & 18 & 18 &  & 2 & 18 & 18 &  & 2 & 18 & 18 \\ 
 & 3 & 38 & 34 &  & 3 & 38 & 34 &  & 3 & 38 & 34 \\ 
 & 4 & 66 & 21 &  & 4 & 66 & 21 &  & 4 & 66 & 21 \\ 
92B & 1 & 6 & 6 & 92C & 1 & 6 & 6 & 92D & 1 & 6 & 6 \\ 
 & 2 & 18 & 18 &  & 2 & 18 & 18 &  & 2 & 18 & 18 \\ 
 & 3 & 38 & 38 &  & 3 & 38 & 38 &  & 3 & 38 & 38 \\ 
 & 4 & 66 & 27 &  & 4 & 66 & 27 &  & 4 & 66 & 27 \\ 
 & 5 & 102 & 2 &  & 5 & 102 & 2 &  & 5 & 102 & 2 \\ 
\hline\hline
\end{tabular}
  \label{tab:clus3d_neighbor}
\end{table} 
%
The parameters of Table~\ref{tab:clus3d} include  the lattice vectors $(\vec{a}_1 , \vec{a}_2 , \vec{a}_3)$,
the lattice ``imperfection'' (IMP)~\cite{PhysRevB.72.060411}, the 
cubicity (C) and the number of completed shells in the cluster (S). 
To understand the meaning of imperfection, one should understand the perfection first. 
The perfection of a cluster measures the completeness of each neighbor shell (Betts shell) as compared 
to the infinite lattice (cf. Table~\ref{tab:clus3d_neighbor}). Accordingly,  a perfect cluster has all neighbor 
shells up to the k-th shell complete, the k-th shell is incomplete, and all shells k+1 and higher are empty. 
The cluster imperfection is defined as the number of sites missing on the (k-1)th and lower shells plus the number of 
sites occupied in shells (k+1)th and higher, IMP=$\sum^{k-1}_{i=1} |N_i- N^{complete}_i| + \sum^{\infty}_{i=k+1}  N_i$, 
where $N_i$ is the number of neighbors in the $i$th shell~\cite{PhysRevB.72.060411}.  
Following such criteria, clusters of N$_c=12$ and N$_c=38$ in Table \ref{tab:clus3d} are considered to be perfect, 
with N$_c$ = 12 the first cluster with a complete nearest-neighbor shell while N$_c$ = 38 is the first 
with a complete next-nearest-neighbor shell (cf. Tables~\ref{tab:clus3d} and~\ref{tab:clus3d_neighbor}).   
The next parameter in Table~\ref{tab:clus3d} is the cubicity~\cite{betts-3d}. It is defined as  
$C=\max(c_1,c_1^{-1})\times \max(c_2,c_2^{-1})$, where $c_1 = 3^{1/2}l/d$ and $c_2 = 2^{1/2}l/f$ are cluster 
parameters defined by the geometric mean of the lengths of the four body diagonals of the cluster, 
$d=\left(d_1d_2d_3d_4\right)^{1/4}$, the six-face diagonals, $f = (f_1f_2f_3f_4f_5f_6)^{1/6}$, and 
the edges, $l = (l_1l_2l_3)^{1/3}$~\cite{PhysRevB.72.060411}. $C=1$ is for a perfect cube, and $C>1$ otherwise. 
A deviation from the cubicity of a perfect cube is a measure of the cubic imperfection. 
Finally, in the last column of 
Table~\ref{tab:clus3d}, we show the number of completed shells on the cluster, for which the 
number of neighbors in that shell on the lattice (LS) and on the cluster (BS) is the same.

In our computations, we utilized only bipartite Betts clusters with small imperfection and good cubicity,
including 10B, 12B, 14B, 16B, 38B, 44B, 80B, 92B, etc.  In finite size scaling, these clusters behave very regularly 
when compared to clusters with large imperfection and/or cubicity.  For example, the choice of such good 
clusters is important in the study of the antiferromagnetic phase diagram of the three-dimensional (3D) 
Hubbard model at half-filling~\cite{PhysRevB.72.060411} and in the zero-temperature properties of quantum 
spin models~\cite{betts-3d}. Furthermore, we consider bipartite clusters because they contain the wavenumber 
$Q=(\pi,\pi,\pi)$.  This allows us to impose the additional particle-hole symmetry on the cluster spectra 
$\rho(K,\omega)=\rho(Q-K,-\omega)$ which reduces the noise in our statistical sampling procedure.  

\section{Hybridization Function Behavior}
In our analysis of the critical behavior of the ALT, we use the local typical density of states TDOS\,($\omega,R = 0$) 
as an order parameter for the description of electron localization (see Fig. 1 of main text). However, the 
imaginary part of the local typical hybridization function 
$\textnormal{Im}\Gamma_{typ}(\omega)=\frac{1}{N_c}\sum_{K=1}^{N_c}\textnormal{Im}[\Gamma_{typ}(K,\omega)]$  
exhibits similar behavior as a function of disorder strength $W$. I.e., the typical hybridization rate between the 
impurity/cluster and the host also vanishes at the transition just as the TDOS. 
\medskip{}
\begin{figure}[th!]
 \includegraphics[trim = 0mm 0mm 0mm 0mm,width=1\columnwidth,clip=true]{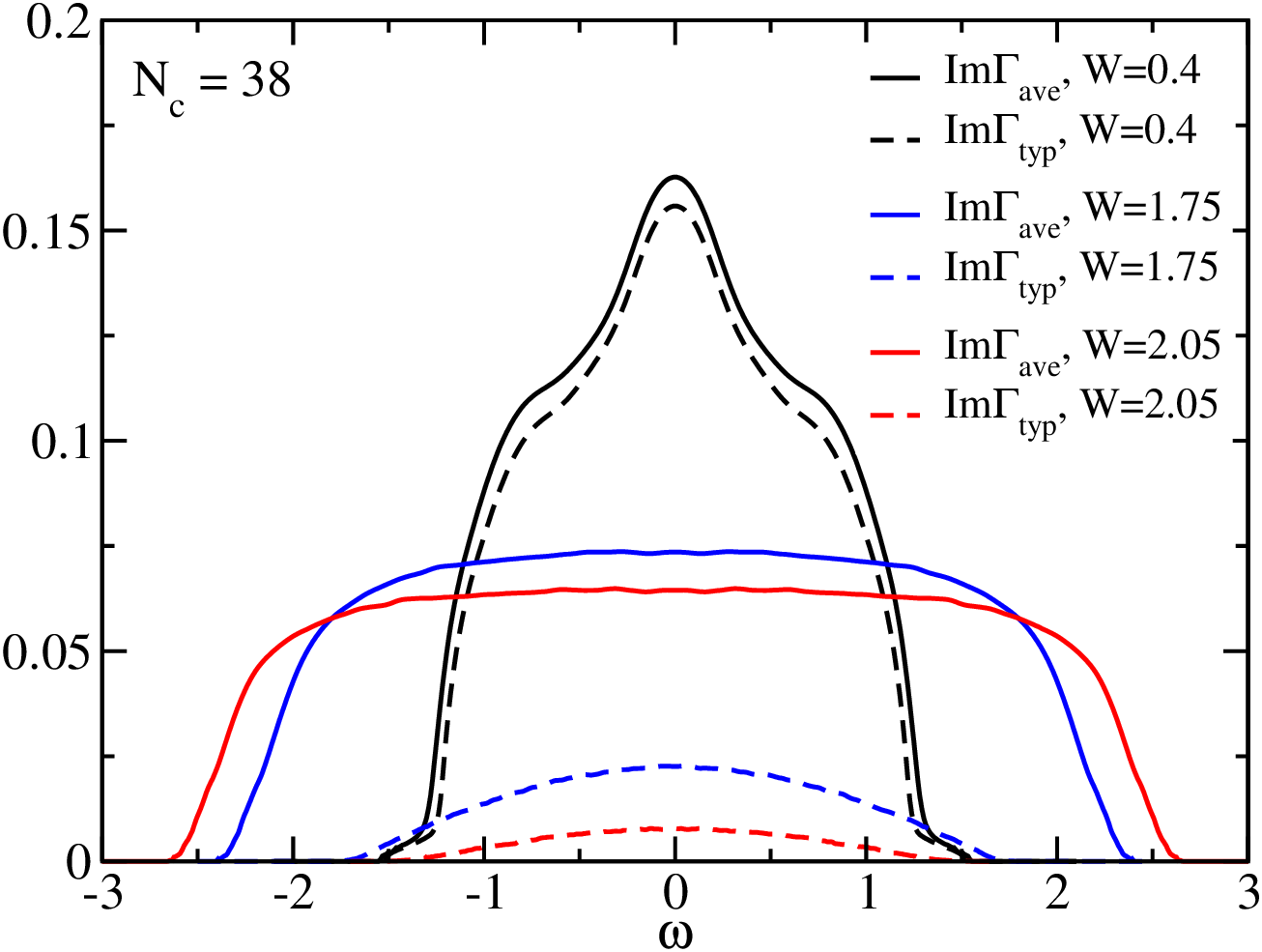}
\caption{(Color online). The imaginary part of local average hybridization function, 
$\textnormal{Im} \Gamma_{ave}(\omega)$, and the typical hybridization function, 
$\textnormal{Im} \Gamma_{typ}(\omega)$, for various disorder 
strengths $W=0.4, 1.75, 2.05$ for cluster size $N_c=38$.}
\label{Fig:Sup_Mat_Fig_Hybr_function}
\end{figure} 

To demonstrate this, we present in Fig.~\ref{Fig:Sup_Mat_Fig_Hybr_function} our results for the imaginary 
part of the local average hybridization function, $\textnormal{Im}\Gamma_{ave}(\omega)$, and the typical 
hybridization function, $\textnormal{Im}\Gamma_{typ}(\omega)$, for N${}_c = 38$ at disorder strengths 
$W = 0.4, 1.75, 2.05$.  Note that the average $\textnormal{Im} \Gamma_{ave}(\omega)$ is calculated using 
the DCA, where the ADOS is used in the self consistency, while $\textnormal{Im} \Gamma_{typ}(\omega)$ is 
evaluated within the TMDCA. As can be seen, at small disorder strength, e.g., $W = 0.4$, our TMDCA and the DCA results 
are numerically equivalent. While the difference between the two procedures become more significant as 
disorder strength increases. Just as in the TDOS from Fig. 1 of the main text, the hybridization rate  
decreases dramatically with disorder, and vanishes at the transition $W_c\approx 2.1$. Also, notice that 
the boundaries of the $\textnormal{Im}\Gamma_{typ}(\omega)$ exhibit re-entrance behavior in the same way 
as the mobility edge trajectories of the TDOS($\omega$,R=0) (cf.~Fig.~\ref{Fig:Sup_Mat_Fig_Hybr_function}). 
Since both the TDOS\,($\omega$,R=0) and $\textnormal{Im}\Gamma_{typ}(\omega)$ go to zero at the same point, 
either of them can be used as an order parameters within the  TMDCA for detecting the Anderson transition. 

\section{Explicit Pole Procedure}
Here, we present in detail how to deal with the poles that emerge on the real-frequency axis close to the 
critical disorder strength, and how to treat the hybridization functions that vanish at different values
of the disorder strength for different coarse graining cells, i.e., cell-selective Anderson localization. 

When $\textnormal{Im}\Gamma_{typ} (K,\omega)$ vanishes, the imaginary part of the cluster-excluded Green function, 
${\cal G}(K,\omega)$, becomes a delta function. 
This can be seen from,
\begin{eqnarray} \label{eqn:cluster_pole}
{\cal G}(K,\omega)
&=&  (\omega - \Gamma_{typ} (K,\omega) - \overline{\epsilon}(K))^{-1} \\
& \Longrightarrow & \textnormal{P} (\omega-\omega^\prime)^{-1} -i\pi \delta(\omega -\omega^\prime ) \nonumber,
\end{eqnarray} 
where $\omega^\prime =  \overline{\epsilon}(K)  + \textnormal{Re} \Gamma_{typ} (K,\omega)  $ and ``P'' denotes 
the principle value.  Obviously, the delta function in Eq.~\ref{eqn:cluster_pole} cannot be represented 
in the conventional way as a list of frequencies on the computer with finite frequency 
resolution $d\omega$.  To avoid this difficulty, we treat those K-cells with such a pole 
with what we call the explicit ``pole-procedure''. This involves replacing such ${\cal G}(K,\omega)$ in 
Eq.~\ref{eqn:cluster_pole} by
\begin{equation}
{\cal G}(K,\omega) = \left\{ \begin{array}{r@{\quad:\quad}l}
-i\pi/{d\omega} &  \omega=\omega^\prime\\
\frac{1}{\omega-\omega^\prime} & \omega \neq \omega^\prime.
\end{array} \right.
\end{equation}
Using this procedure, the singular behavior of ${\cal G}(K,\omega)$ can be properly captured. 
The difficulty is that as we approach W$_c$ for a given N$_c$, $\Gamma_{typ} (K,\omega)$ 
for individual cells goes to zero at different rates -- a manifestation of the cell-selective Anderson 
localization.  So we have to determine which of these cells we need to apply the ``pole-procedure'' to. 
To address this problem we apply the procedure to a cell when $(-1/\pi)\times\textnormal{Im}\Gamma_{typ} (K,\omega^\prime)$ 
$<$ $a\times d\omega^\prime$, here $a \agt 1$ is a parameter which measures the minimum number of 
pixels required to represent a pole approaching the real frequency axis. Our numerical experience demonstrate 
that such a criterion works nicely while spurious results are obtained otherwise.

\section{Critical Parameters}
The critical disorder strength $W_c$ are reported in Table~\ref{Tab:critical_nu} 
for various cluster sizes. $W_c$ was determined as
the $W$ where the TDOS\,($\omega=0$) vanishes. Observe that as N${}_c$ increases, $W_c$ systematically 
increases with $W_c^{N_c\geq12}\approx 2.10 \pm 0.01$, showing a quick convergence with N$_c$. 


\begin{table}[h] 
\centering
\caption{The calculated critical disorder strength $W_c$ for various cluster sizes.  $W_c$ is defined 
as the vanishing of the TDOS\,($\omega=0$).}
\begin{tabular}{ccccccc}
\hline\hline
N$_c$ & & & W$_c$ & & &   \\ 
\hline
1 & & & 1.66$\pm$0.01 & & &  \\ 
6 & & & 1.68$\pm$0.01  & & &  \\ 
10 & & & 1.90$\pm$0.01  & & &   \\ 
12 & & & 2.10$\pm$0.01  & & &   \\ 
14 & & & 2.10$\pm$0.01  & & &   \\ 
16 & & & 2.10$\pm$0.01  & & &  \\ 
38 & & & 2.10$\pm$0.01  & & &  \\ 
44 & & & 2.10$\pm$0.01  & & &   \\ 
80 & & & 2.10$\pm$0.01  & & &   \\ 
92 & & & 2.10$\pm$0.01  & & &  \\ 
\hline\hline
\end{tabular}
\label{Tab:critical_nu}
\end{table}

\section{Avoiding Self-Averaging}
The averaging procedure used to calculate the typical spectra is not unique.  As noted in the main text, 
while CTMT~\cite{Ekuma-Arxiv} works well for one and two dimensions, in three dimensions it suffers 
from effective self-averaging for large clusters. This is due to the fact that close to the criticality, 
there is a mixture of localized and extended states above and below the localization edge given by the
TDOS. These energy scales need to be treated differently. The CTMT fails to do this as can be seen by 
inspecting the spectral density used in the CTMT self-consistency
\begin{equation} \label{rho_typ}
\rho_{typ}^c(K,\omega)
=
\exp \left\langle \ln \rho^c (K,\omega,V_i) \right\rangle. 
\end{equation}
In forming the Fourier transform
\begin{equation} \label{rho_FT}
\rho^c (K,\omega,V_i) = \sum_{X,X'} \exp(iK\cdot(X-X'))\rho^c (X,X',\omega,V_i)
\end{equation}
we average over the cluster coordinates $X$ and $X'$, including the local part, $X=X'$.  I.e., the local 
DOS is first averaged over the cluster sites and then Fourier transformed to form the local part of $\rho^c(K,\omega)$.  
So for large clusters our procedure reduces to linear averaging of the local part instead of geometrical 
averaging.  Thus, the host Green function constructed from $\rho_{typ}^c(K,\omega)$ is unaware of the 
TDOS and thus is unable to distinguish between the energies above and below the localization edge.  As 
demonstrated in the main text, to avoid such self-averaging in the TDOS, we propose the Typical Medium DCA 
(TMDCA) method. Here, the cluster-momentum-resolved typical density of states (TDOS) for each $K$ is split 
into local and nonlocal parts. The local part is treated with geometrical averaging over disorder configurations, 
while the non-local part is treated either with an algebraic or geometric averaging over the disorder 
configuration. 

To do this, we have utilized two schemes. The first scheme is what we call \textit{linear-log} procedure which 
is what we utilized in the main text.  Here, we treat the local part with a geometrical averaging while the 
non-local part is approximated algebraically using linear averaging as
\begin{multline} \label{Eq:log-linear}
\rho_{typ}^c(K,\omega)
=
\exp\left(\frac{1}{N_c} \sum_{i=1}^{N_c} \left\langle \ln \rho_{i}^c (\omega,V_i)  \right\rangle\right) \times \\ 
\left\langle \frac{\rho^c(K,\omega,V_i)}{\frac{1}{N_c} \sum_{i} \rho_{i}^c (\omega,V_i)} \right\rangle.
\end{multline}
The second scheme is what we call the \textit{log-log} procedure which again involves the treatment of 
the local part with geometrical averaging and the non-local part is also treated with a log averaging as 
\begin{multline} \label{Eq:log-log}
\rho_{typ}^c(K,\omega)
=
\exp\left(\frac{1}{N_c} \sum_{i=1}^{N_c} \left\langle \ln \rho_{i}^c (\omega,V_i)  \right\rangle \right) \times \\ 
\exp \left( \left\langle \ln \frac{\rho^c(K,\omega,V_i)}{\frac{1}{N_c} \sum_{i} \rho_{i}^c (\omega,V_i)} \right\rangle \right).
\end{multline}

\medskip{}
\begin{figure}[th!]
 \includegraphics[trim = 0mm 0mm 0mm 0mm,width=1\columnwidth,clip=true]{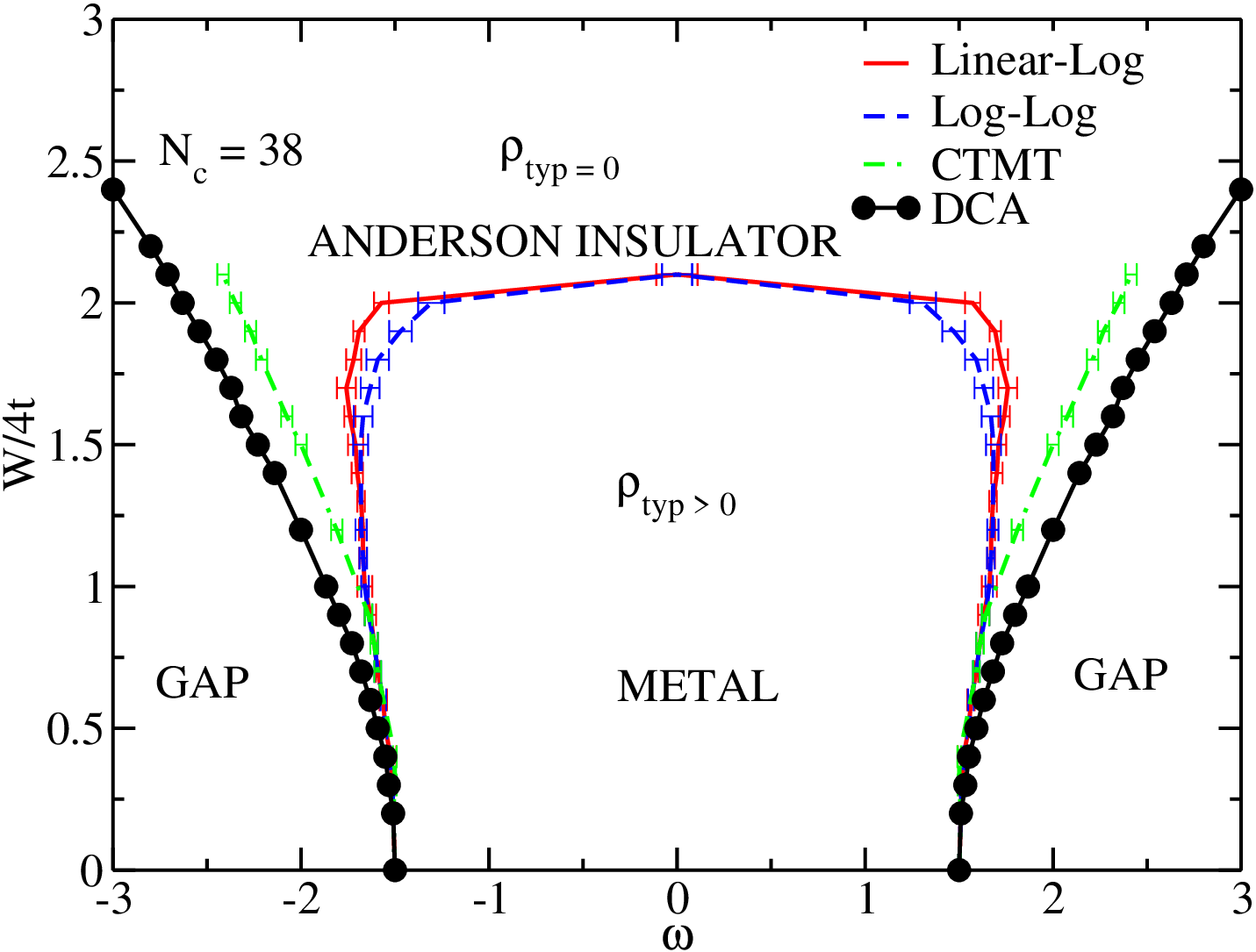}
\caption{(Color online). A comparison of the phase diagram of the Anderson localization transition in 3D 
obtained  from cluster approximations with N${}_c=38$ using  CTMT and the TMDCA (linear-log 
and log-log) schemes. Observe that in the CTMT, as consequence of self-averaging, the higher disorder 
behaviors which are captured in our TMDCA are totally missed and the critical disorder strength is also 
severely over-estimated.}
\label{Fig:Phase-Diagram_Nc38-compare.eps}
\end{figure} 

We note most importantly that while there are different behaviors of the two methods around the re-entrance 
region, both approaches systematically converge to the same critical disorder strength 
$W_c^{N_c\geq12}\approx 2.1\pm 0.01$. However, the \textit{linear-log} procedure is generally more robust 
than the \textit{log-log} method. The latter displays a slower convergence around the re-entrance region, 
requiring far larger clusters before the re-entrance region converges to the numerical experimental results. 
It also may not be adequate to study  localization phenomena in real materials, since it
is not clear how one would perform a geometric average of the band off-diagonal components of the spectral
density, since they are not positive definite. The comparison of the phase diagram obtained using CTMT and 
the TMDCA: log-log and linear-log formalisms 
is depicted in Fig~\ref{Fig:Phase-Diagram_Nc38-compare.eps}. As it is evident from 
Fig~\ref{Fig:Phase-Diagram_Nc38-compare.eps}, the two new schemes converge to the same 
critical disorder strength but behave differently around the re-entrance region while the CTMT will eventually 
converge to a disorder strength far greater than W$_c$. We further remark that the re-entrance trajectory of the 
mobility edge is totally missed in the CTMT as a consequence of self-averaging in the cluster. 

We note that in both schemes, at small N${}_c$, $\approx$ 100 self-consistent iterations are required 
to achieve a convergence, while for relatively large N${}_c$, far fewer iterations are required. The convergence 
criterion in both limits is achieved when the TDOS\,($\omega = 0$) does not fluctuate anymore with iteration number 
within the error bars. 

Finally, we note that many other definitions of the typical medium which avoid self averaging are possible, 
including the use of only the local part of Eq.~\ref{Eq:log-linear}, i.e.,
\begin{equation}
\rho_{typ}^c(K,\omega)
=
\exp\left(\frac{1}{N_c} \sum_{i=1}^{N_c} \left\langle \ln \rho_{i}^c (\omega,V_i)  \right\rangle\right) \,.
\end{equation}
However, this method was rejected since it does not meet all of the the criteria discussed in the main 
text.  In this case, this formalism does not recover the DCA in the weak coupling limit.